\documentclass[a4paper, preprint, aip,nobibnotes]{revtex4-1}

\usepackage{graphicx}
\usepackage{dcolumn}
\usepackage{amsmath}
\usepackage{textcomp}
\usepackage[usenames,dvipsnames]{color}

\begin{document}
\newcommand{\revEdit}[1]{\textcolor{black}{#1}}

\title{Fabrication of quantum dots in undoped Si/Si$_{0.8}$Ge$_{0.2}$ heterostructures using a single metal-gate layer}

\author{T. M. Lu}
\email{tlu@sandia.gov}
\affiliation{Sandia National Laboratories, Albuquerque, New Mexico 87185, USA}
\author{J. K. Gamble}
\affiliation{Sandia National Laboratories, Albuquerque, New Mexico 87185, USA}
\author{R. P. Muller}
\affiliation{Sandia National Laboratories, Albuquerque, New Mexico 87185, USA}
\author{E. Nielsen}
\affiliation{Sandia National Laboratories, Albuquerque, New Mexico 87185, USA}
\author{D. Bethke}
\affiliation{Sandia National Laboratories, Albuquerque, New Mexico 87185, USA}
\author{G. A. Ten Eyck}
\affiliation{Sandia National Laboratories, Albuquerque, New Mexico 87185, USA}
\author{T. Pluym}
\affiliation{Sandia National Laboratories, Albuquerque, New Mexico 87185, USA}
\author{J. R. Wendt}
\affiliation{Sandia National Laboratories, Albuquerque, New Mexico 87185, USA}
\author{J. Dominguez}
\affiliation{Sandia National Laboratories, Albuquerque, New Mexico 87185, USA}
\author{M. P. Lilly}
\affiliation{Sandia National Laboratories, Albuquerque, New Mexico 87185, USA}
\author{M. S. Carroll}
\affiliation{Sandia National Laboratories, Albuquerque, New Mexico 87185, USA}
\author{M. C. Wanke}
\affiliation{Sandia National Laboratories, Albuquerque, New Mexico 87185, USA}

\date{\today}

\begin{abstract}
Enhancement-mode Si/SiGe electron quantum dots have been pursued extensively by many groups for \revEdit{their} potential in quantum computing.  Most of the reported dot designs utilize multiple metal-gate layers and use Si/SiGe heterostructures with Ge concentration close to 30\%.  Here we report the fabrication and low-temperature characterization of quantum dots in Si/Si$_{0.8}$Ge$_{0.2}$ heterostructures using only one metal-gate layer.  We find that the threshold voltage of a channel narrower than 1 $\mu$m increases as the width decreases.  The higher threshold can be attributed to the combination of quantum confinement and disorder.  We also find that the lower Ge ratio used here leads to a narrower operational gate bias range. The higher threshold combined with the limited gate bias range constrains the device design of lithographic quantum dots.  We incorporate such considerations in our device design and demonstrate a quantum dot that can be tuned from a single dot to a double dot.  The device uses only a single metal-gate layer, greatly simplifying device design and fabrication. 
\end{abstract}

\maketitle

Electron spins are one of the most promising candidates for implementing solid-state qubits\cite{Loss}.  Si, in particular, is a widely pursued host material for spin qubits, due to its long coherence times\cite{Tyryshkin2003} resulting from its weak spin-orbit interaction and the possibility of obtaining isotopically pure $^{28}$Si which has zero nuclear spin\cite{Zwanenburg2013}.  In Si, manipulation of individual electron spins can be achieved in electrostatically defined quantum dots in addition to donor-bound electrons as proposed by Kane\cite{Kane1998}.  A typical starting platform for making quantum dots is a two-dimensional electron gas (2DEG), which either resides at the Si/SiO$_2$ interface in the case of Si metal-oxide-semiconductor field-effect transistors (MOSFETs), or in a strained Si quantum well in the case of Si/SiGe heterostructures.  A Si/SiGe heterostructure has epitaxial interfaces, which \revEdit{are} much less disordered than amorphous SiO$_2$ on Si and \revEdit{result} in much higher electron mobilities.  For Si/SiGe heterostructures, one can either use modulation-doping\cite{Schaffler1997} and fabricate gates to locally deplete electrons, or employ a MOSFET-like enhancement-mode architecture\cite{Lu2007} and induce electrons only where desired.  Due to higher starting mobility and better device stability, the enhancement-mode architecture has gained popularity in recent years. Device designs, dot operation, and qubit manipulation have all been reported using this enhancement-mode architecture\cite{q-dot1,kim2014,eng2015,Petta2015}.

In most of the reported dot designs, the starting Si/SiGe heterostructure has a Ge concentration close to 30\%, and the gates used to define quantum dots are structured in multiple metal layers.  There has not been much discussion on the choice of \revEdit{this} combination.  \revEdit{Since electron mobility is the most obvious and convenient gauge for material quality and the record electron mobility\cite{Lu2009} was demonstrated in a heterostructure with a Ge content of 20\%, we chose the same composition for the relaxed SiGe buffer layers as we started fabricating Si/SiGe quantum dots.  The higher achievable mobility was expected to reduce the impact of disorder.}  At first glance, the main difference between \revEdit{Si/Si$_{0.7}$Ge$_{0.3}$ and Si/Si$_{0.8}$Ge$_{0.2}$} is simply the barrier height in the 2DEG normal direction.  This difference in barrier height is naively not expected to mandate a change in quantum dot design, since most design considerations center around creating lateral confinement of electrons to form quantum dots.  \revEdit{Surprisingly, we learned that a reduced Ge concentration affects device performance indirectly but strongly.}  In this work we study the physical mechanisms of this effect through low-temperature experiments and simulations, and discuss the constraints imposed on device design.  \revEdit{Accounting for the constraints}, we demonstrate \revEdit{electrical control of} quantum dots in a Si/Si$_{0.8}$Ge$_{0.2}$ heterostructure.  %The transport characteristics of a fabricated quantum dot can be tuned from single-dot-like to double-dot-like.   
Furthermore, the fabrication process flow presented here uses only a single metal-gate layer, which greatly simplifies device fabrication, shortens device turnaround times, and improves yield by avoiding potentially leaky metal-insulator-metal stacks. 

\revEdit{In this work we used} undoped Si/Si$_{0.8}$Ge$_{0.2}$ heterostructures grown by Lawrence Semiconductor Research Laboratory.  
%Three types of virtual substrates were used: 150-mm (100) 100-$\Omega\cdot$cm p-type Si wafers, 150-mm (100) 0.001-$\Omega\cdot$cm p-type Si wafers, and 150-mm (100) silicon-on-insulator (SOI) wafers.  \revEdit{For this work}, the three types of starting substrates \revEdit{are} irrelevant.  \revEdit{For each SOI wafer, both the handle and the device layer were p-type with resistivities of 0.001-$\Omega\cdot$cm and 1000-$\Omega\cdot$cm, respectively.}  
\revEdit{A virtual substrate was} made by growing a 2-$\mu$m linearly graded buffer layer on a Si wafer at a grading rate of 10\%/$\mu$m, followed by a 1-$\mu$m Si$_{0.8}$Ge$_{0.2}$ relaxed buffer layer.  Chemical mechanical polishing (CMP) \revEdit{removed} 150 nm of the relaxed buffer layer to reduce surface roughness.  After CMP, epi-layers of 400 nm Si$_{0.8}$Ge$_{0.2}$, 20 nm Si, 35 nm Si$_{0.8}$Ge$_{0.2}$, and 3 nm Si were grown in order.  
The heterostructures were processed in the Si foundry at Sandia National Laboratories.  A similar process flow for SiGe heterostructures has been reported earlier\cite{q-dot1}.  
%Ion implantation of As at 60 KeV with a fluence of 1$\times$10$^{16}$ cm$^{-2}$ and thermal activation \revEdit{formed} highly doped regions in order to make subsequent ohmic contact.  A SiO$_2$ layer at least 100-nm thick was deposited \revEdit{using either} high-density-plasma chemical-vapor-deposition (CVD) or low-pressure CVD.  Via holes were etched to provide electrical access to the highly doped Si region, and Ti/TiN/W/TiN was deposited to create ohmic contacts.  \revEdit{At this point the wafers were diced into $\sim$1$\times$1 cm$^2$ pieces before the remaining die-level processing steps.}  The CVD SiO$_2$ layer was removed in a 100$\times$100 $\mu$m$^2$ device window by HF.  
Electron-beam lithography (EBL) \revEdit{defined} the nanostructured gate patterns.  15 nm Al$_2$O$_3$ and 100 nm Al were then deposited in the same run, followed by lift-off.  The Al$_2$O$_3$ layer that insulates the gate \revEdit{from} the substrate was formed by depositing Al in the presence of O$_2$.

We made a series of narrow transistors with variable channel widths, ranging from 100 nm to 5 $\mu$m \revEdit{wide}. Each narrow gate was 2 $\mu$m long. At each end of a narrow gated region, the enhancement gate was flared quickly to a width exceeding 10 $\mu$m and was extended to overlap two high doped regions on each \revEdit{end} of the narrow wire to make two \revEdit{ohmic} contacts on each \revEdit{end}.  The device threshold voltages were characterized at 4 K by monitoring the onset of channel current at a quasi-dc drain-source bias ($V_{DS}$) \revEdit{of 1 mV} against the gate voltage ($V_G$).  Using the four ohmic contacts for each device, we extracted two threshold voltages, one for the narrow channel, and one for the wide channel.  The difference in the two threshold voltages for each device was then tabulated.  Some \revEdit{narrow} channels did not turn on \revEdit{below} $V_G=2$ V, the highest voltage applied.  These devices were excluded from the calculation of threshold shifts.  \revEdit{Fig.~1~(c) shows the average shift and the standard deviation of threshold voltages of the narrow channels, and Fig.~1~(d) shows the fraction of working devices.  The non-uniform standard deviation arises from outliers that had a threshold voltage much higher than the other devices in the same group.}

\revEdit{It is clear from Fig.~1~(c) and (d) that the threshold voltage of a narrow channel shifts higher and a larger fraction of devices do not turn on  as the channel gets narrower.}  In the case of 100-nm-wide channels, none of the devices showed turn-on below V$_G=2$ V.  This is in stark contrast to 100-nm channel widths routinely achieved using similar structures with a 30\% Ge concentration\cite{eng2015,Petta2015}.  \revEdit{We attribute the non-working channels and the threshold shifts to the lower Ge concentration used in this work in combination with two other mechanisms, lateral quantum confinement and disorder.}

To understand the lateral quantum confinement effect, we performed simulations to analyze the nanowire geometry explored by the experiment.  For each wire width, we computed the electrostatic landscape by solving Poisson's equation in 2D with the finite element method in COMSOL Multiphysics using the geometry shown in Fig.~1~(a).  We then used this potential energy landscape as input to Schr\"odinger's equation, \revEdit{and solved} for the energies of the 1D nanowire sub-bands.  Since different wire widths produce different 2D ground state energies for \revEdit{the same applied voltage,} the lowest sub-band begins filling at different applied voltages for different wire widths.

To check if this model is consistent with our experiment, we note that the ground state energy of the 2D system is linear with gate voltage, as shown by the computational results in Fig.~1~(b). Hence, we may write
\begin{equation}
E_0^{L} = m_L V_G,
\end{equation}
where $E_0^{L}$ is the ground state energy of the wire of width $L$, $m_L$ is the slope (determined computationally), and $V_G$ is the applied gate voltage.
By inverting this relationship, we can express the voltage shift of  a narrow wire compared to an infinitely wide wire as,
\begin{equation}
\Delta V_L =\frac{ E_0}{m_L m_\infty} \left( m_\infty - m_L \right),
\end{equation}
where $m_\infty$ is the slope in the infinite wire-width limit, \revEdit{and $E_0$ represents a constant threshold energy independent of wire width. We assume the threshold occurs when the ground state energy level of a specific wire, $E_0^{L}$, reaches $E_0$.}
Noting that the values of $m_L$ saturate for the wide wires we considered, we take $m_\infty \approx m_{5000} = -930$~meV/V. 
\revEdit{Treating} $E_0$ as a free parameter and \revEdit{fitting} the model to the data using a chi-squared test statistic, \revEdit{we obtained} a goodness of fit $p=0.9953$,
indicating excellent consistency with experiment.

In addition to quantum confinement, we believe that disorder also plays an important role in the observed threshold voltage shift.  In the presence of disorder, a percolation conduction threshold is associated with each narrow channel.  The local conduction threshold may be higher for narrow channels if the widths of the channels are smaller than the characteristic length scale of the disorder potential.  Statistical variation mandates that some devices would experience stronger disorder than others and show higher threshold voltages.  The spread of the \revEdit{experimentally observed} threshold voltage shifts for a fixed channel width give support to the effect.  

The two mechanisms discussed above lead to a higher average threshold voltage as well as a wider spread.  This is true for all Si-based narrow channels.  For Si MOSFETs, a shift in threshold voltage is not a critical problem.  In principle, one can always increase the gate voltage to induce more electrons, and eventually the conduction threshold is overcome at high enough densities.  {\it \revEdit{Unfortunately} for Si/SiGe heterostructure FETs, there exists an upper density limit\cite{TMLU-tunnel}}.  For shallow Si quantum well channels this maximum density signals the onset of population of a surface channel at the oxide/semiconductor interface.  For deep Si quantum wells, a non-equilibrium distribution of electrons collect in the well, with a maximum density controlled by tunneling to a parallel surface channel.  \revEdit{A detailed discussion of this non-equilibrium charge distribution can be found in Ref.~\onlinecite{TMLU-tunnel}.}  In both cases, the saturation density is dependent on the barrier height, or in turn, the Ge concentration of the SiGe layers\revEdit{, with the maximum density independent of the channel depth in the latter case\cite{TMLU-tunnel}.}  \revEdit{The wells used here are considered deep wells and as such,} the electron density can be higher than the thermal equilibrium value and is dynamically limited by slow tunneling.  Experimentally this upper density limit is $\sim$3$\times$10$^{11}$ cm$^{-2}$ for Si$_{0.8}$Ge$_{0.2}$ barriers and $\sim$7$\times$10$^{11}$ cm$^{-2}$ for Si$_{0.7}$Ge$_{0.3}$ barriers\cite{TMLU-tunnel}.  We obtained a similar upper density limit in our structures using Hall bar testers fabricated from the same material.  This density limit, nonlinear in barrier height, limits the tunable density range in Si/Si$_{0.8}$Ge$_{0.2}$ to less than half the density range in Si/Si$_{0.7}$Ge$_{0.3}$.  The maximum density \revEdit{results in} a maximum Fermi energy. For gate voltages beyond the density saturation bias, a possibly more serious effect occurs, namely, the oxide/semiconductor interface forms a second \revEdit{accumulation layer} and eventually accumulates a high enough electron density to become conducting\cite{Sturm-noeq}. When the surface channel conduction occurs, this surface channel screens the electric fields to the buried Si quantum well channels, and reduces the electron density in the Si quantum well to its thermal equilibrium value.  The thermal equilibrium value could be below the conduction threshold, such that the channel conductivity drops to zero. This appears to happen in all our samples, in which we observe a sudden drop in current to zero at a gate bias beyond threshold, \revEdit{as shown in the inset of Fig.~1~(d)}.

Since the current in our devices has to flow serially through both the wide and narrow channel areas, the maximal Fermi energy and the sudden turnoff \revEdit{pose} a serious problem if the maximum Fermi energy in the wide gated channel area is not much higher or even lower than the conduction threshold energy of the narrow channel. In the former case, the narrow channel barely turns on before the wide channel turns off, and hence the resistance is very high. In the latter case, when the lateral confinement lifts the threshold energy level significantly, the narrow wire will not reach threshold until the gate is biased beyond the turn-off voltage of the wide channel area. Thus for \revEdit{the} narrowest wires there will never be a \revEdit{simultaneous} conduction path through both regions.  Using Si/Si$_{0.8}$Ge$_{0.2}$ heterostructures instead of Si/Si$_{0.7}$Ge$_{0.3}$ thus severely constrains the operational window of $V_G$.  A multiple-metal-layer architecture with independent controls for the reservoirs and nanostructures circumvents this density limit problem since the potential landscape can be locally tuned.

Having understood the major difference between Si/Si$_{0.7}$Ge$_{0.3}$ and Si/Si$_{0.8}$Ge$_{0.2}$, we now turn to our results of fabricating quantum dots using Si/Si$_{0.8}$Ge$_{0.2}$ heterostructures.  Since the main difficulty in using lower Ge concentration is the high threshold voltages for narrow channels, the quantum dot designs have to be enlarged compared to what are used in Si MOSFETs or in Si/Si$_{0.7}$Ge$_{0.3}$, unless a multiple-metal-layer architecture is adopted.  For this work we chose to use only one metal layer and therefore chose to make large quantum dots.  A larger quantum dot has lower charging energies and requires lower temperatures to resolve transport features.  However, the larger feature sizes are more easily fabricated.  Furthermore, the one-metal-layer architecture has fewer fabrication steps and fewer potentially leaky metal-insulator-metal stacks.  This simpler process flow, together with the relaxed nano-patterning accuracy, significantly shortens fabrication turnaround times and improves device yield.

The starting wafer for the dot work had nominally identical growth parameters as previously described.  Here we did every step at die-level.  Ti/Au alignment marks were first deposited, followed by ion implantation of P at 20 KeV and 75 KeV with a fluence of 5$\times$10$^{14}$ cm$^{-2}$ for each energy.  A rapid thermal anneal at 625~$^\circ$C for 30 sec activated the implanted dopants for ohmic contacts.  A blanket 20-nm-thick Al$_2$O$_3$ layer was deposited in an atomic-layer-deposition system at 200~$^\circ$C.  After etching vias through Al$_2$O$_3$, we deposited a blanket 2 nm Ti and \revEdit{40} nm Au film.  This metal layer was patterned by EBL and etched in an ion mill.  The exposed metal areas were milled away, \revEdit{leaving} nanostructure gates and bond pads simultaneously.  Fig.~2~(a) shows a scanning electron micrograph of a \revEdit{fabricated} quantum-dot device.  Also shown is the circuit setup used for the data presented below.  The names of the gates are labeled in yellow.  
%This design was similar to dot designs demonstrated in Si MOSFETs\cite{Tracy2013}, but was scaled several times larger.  
The narrowest point of the channel had a width of $\sim$200 nm.

The device shown in Fig.~2~(a) was studied in a $^3$He cryostat with a base temperature of $\sim$ 0.3 K at $V_{DS}=$ 10 $\mu$V using standard lock-in techniques.  We focused on the upper channel and varied the voltages for UL, UC, and UR to form quantum dots.  The voltages used for AGU, AGL, LL, LC, and LR were \revEdit{kept} constant at 0.633, -0.35, 0, 0, and 0 V, respectively, for the data presented here.  
In Fig.~2~(b) we show a series of stability plots against the UL and UR gates at different UC voltages.  At low (less negative) UC, single-quantum-dot behavior is dominant in the lower left corner of a UR vs. UL plot, with approximately equal couplings to the UL and UR gates.  Upon making UC more negative, the dot is broken up into two halves by the electrostatic potential, and forms a pair of quantum dots coupled by tunneling.  The isolated transport peaks \revEdit{observed showed the expected} triple points for tunnel-coupled double-quantum-dots\cite{RMP2002}.  \revEdit{From stability diagrams, we extract the following gate-dot capacitances for the double-dot system: $C_{UL,L}=$ 4.0 aF, $C_{UC,L}=$ 5.2 aF, $C_{UR,L}=$ 1.2 aF, $C_{AGU,L}=$ 63 aF, $C_{AGL,L}=$ 13 aF, $C_{LL,L}=$ 1.3 aF, $C_{LC,L}=$ 1.4 aF, $C_{LR,L}=$ 0.48 aF, $C_{UL,R}=$ 1.3 aF, $C_{UC,R}=$ 5.2 aF, $C_{UR,R}=$ 4.0 aF, $C_{AGU,R}=$ 79 aF, $C_{AGL,R}=$ 13 aF, $C_{LL,R}=$ 0.63 aF, $C_{LC,R}=$ 1.3 aF, $C_{LR,R}=$ 0.48 aF.  These capacitances are consistent with a lithographic double-dot defined by the gates.}  The inter-dot tunnel barrier is not only tuned by the UC gate, but also by the UL and UR gates, as is evidenced from \revEdit{the} transition from single-dot-like to double-dot-like characteristics moving from the upper right corner to the lower left corner at a fixed UC voltage.  This is a trade-off of using a single-metal-gate-layer design; controls over tunnel barriers and dot occupations are more intertwined together and require more \revEdit{sophisticated} device tuning.

To corroborate our single to double-dot interpretation of Fig.~2~(b), we performed capacitive modeling of our nanostructure using FastCap to obtain a simulated charge stability diagram.
We used voltages identical to our experiment, modeled the electrical leads below AGU as a metallic brick, and the two quantum dots as 10 nm thick conductive bricks with $400\times400$~nm in-plane dimensions, as shown in Fig.~2~(a).
The two dots were separated by a distance $d$, which we varied. 
As shown in Fig.~2~(c), by \revEdit{changing the distance from} $0.1$ to $10$ nm, we are able to achieve a clear transition from a single- to double-dot charge stability signature. 
These plots are qualitatively consistent with those in Fig.~2~(b), indicating that one needs only to consider (plausible) small dot separations when explaining the data.

In summary, we \revEdit{discovered} the constraints imposed by using Si/SiGe heterostructures with a Ge concentration of 20\%, lower than what is used in other reports.  \revEdit{We found that the limited tunable density range prevents conduction through channels gated by a single layer gate which contains both very narrow and wide channel widths.}  With a scaled-up quantum dot design, we demonstrated operation of lithographically defined quantum dots using Si/Si$_{0.8}$Ge$_{0.2}$ heterostructures, and were able to induce a single-dot to double-dot transition.  The presented fabrication process flow requires only one metal-gate-layer for all gates and bond pads, significantly reducing device fabrication turnaround times.  This may find use in situations where high throughput of simple quantum dots is required, such as studying the statistics of quantum dot properties.

This work was performed, in part, at the Center for Integrated Nanotechnologies, a U.S. DOE, Office of Basic Energy Sciences, user facility.  Sandia National Laboratories is a multi program laboratory managed and operated by Sandia Corporation, a wholly owned subsidiary of Lockheed Martin Corporation, for the U.S. DOE's National Nuclear Security Administration under contract DE-AC04-94AL85000.

%\bibliography{SingleLayerSiGeDot}


\newpage
\begin{figure}

\includegraphics[width=3.3 in]{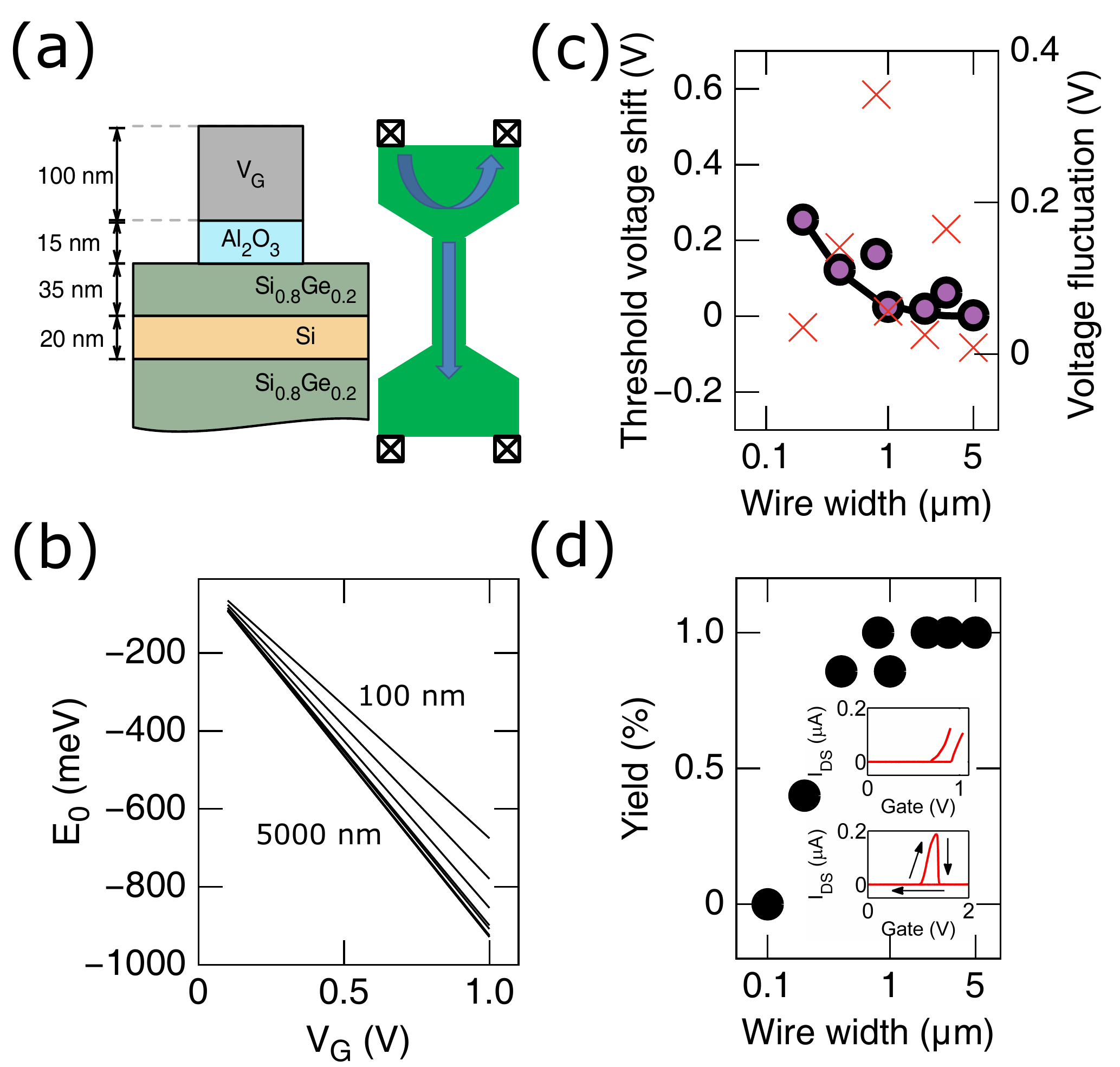}
 \caption{%Variation of threshold voltage with channel width.
 (a): \revEdit{Schematic cross-section and top view of the measured/modeled FET.  The arrows illustrate the current paths of the wide and narrow channels.}
 (b): Calculated ground state energy of the channel produced by applying a voltage on the gate indicated in (a). The lines correspond to the wire widths measured in panel (c).
 (c): \revEdit{Threshold voltage shift (solid dots) and its standard deviation (x markers) as a function of wire width.}  The line is the model fitted to a single-parameter model, obtaining a goodness of fit $p=0.9953$.
 \revEdit{(d): Fraction of functional devices (yield) as a function of wire width.  Top inset: representative turn-on curves at $V_{DS}=$ 1 mV for a 200-nm wide channel (right curve) and an 800-nm wide channel (left curve).  Bottom inset: a representative turn-on curve showing that the channel shuts off beyond a critical gate voltage.}
  \label{fig:wireWidthVar}}
\end{figure}

\begin{figure}

\includegraphics[width=3.3 in]{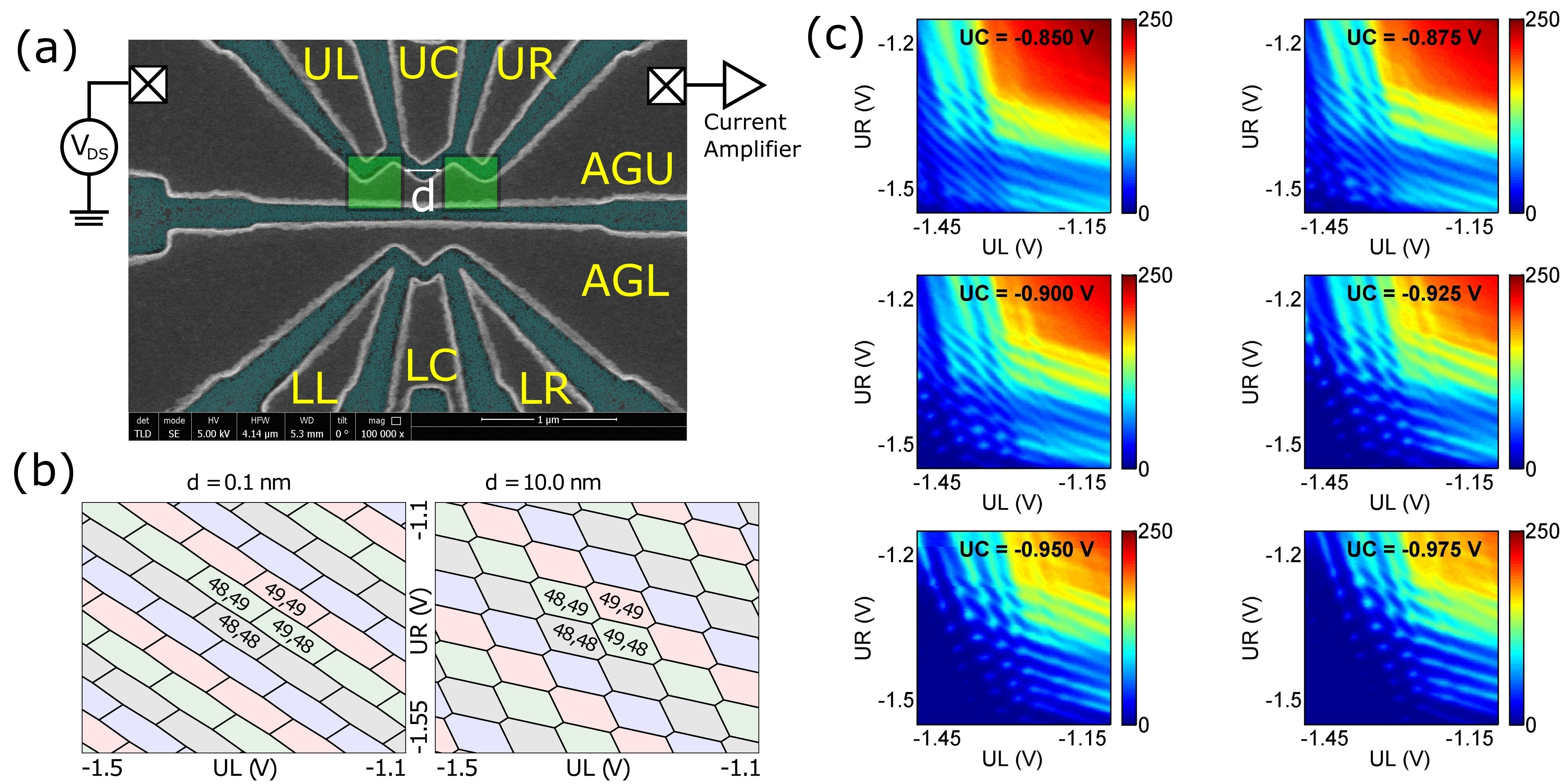}
\caption{\label{deviceSchematic} %The quantum dot device studied in this work.
(a) Scanning electron micrograph of the device. Lighter gray with yellow labels are Ti/Au gates on top of the Si/SiGe substrate.  The cyan pseudo-colored regions are exposed Al$_2$O$_3$.  The capacitance calculations assume two square quantum dots with side length 400 nm as indicated in green. 
These two dots are separated by a distance $d$.
(b) Stability diagrams for the quantum dot in the upper channel at $V_{UC}=$ -0.85, -0.875, -0.9, -0.925, -0.95, and -0.975 V.  The measurement temperature was 0.3 K, and the drain-source bias was 10 $\mu$V.
(c) Capacitance simulations for differing dot separation $d$. We see that even a modest dot separation of 10 nm is sufficient to transition from single- to double-dot
behavior.
These calculations indicate that the observed charge stability diagrams in (b) likely describe a transition between lithographic dots. 
}
\end{figure}

\end{document}